\documentclass[aps,preprint]{revtex4}
\usepackage{eurosym}
\usepackage{amssymb,epsf}
\usepackage{amsmath,amsfonts}
\usepackage{graphicx}
\usepackage{epsfig}
\usepackage{epstopdf}
\usepackage{graphicx}

\begin{document}

\author{N. Farhangkhah \thanks{%
email address: farhangkhah@iaushiraz.ac.ir}}
\title{Topologically nontrivial black holes in Lovelock-Born-Infeld gravity}
\affiliation{Department of Physics, Shiraz Branch, Islamic Azad University, Shiraz 71993,
Iran}

\begin{abstract}
We present the black hole solutions possessing horizon with
nonconstant-curvature and additional scalar restrictions on the base
manifold in Lovelock gravity coupled to Born-Infeld (BI) nonlinear
electrodynamics. The asymptotic and near origin behavior of the metric is
presented and we analyze different behaviors of the singularity. We find
that, in contrast to the case of black hole solutions of BI-Lovelock gravity
with constant curvature horizon and Maxwell-Lovelock gravity with nonconstant
horizon which have only timelike singularities, spacelike, and timelike
singularities may exist for BI-Lovelock black holes with
nonconstant curvature horizon. By calculating the
thermodynamic quantities, we study the effects of nonlinear electrodynamics
via the Born-Infeld action. Stability analysis shows that black holes with
positive sectional curvature, $\kappa$, possess an intermediate unstable
phase and large and small black holes are stable. We see that while Ricci
flat Lovelock-Born-Infeld black holes having exotic horizons are stable in
the presence of Maxwell field or either Born Infeld field with large born
Infeld parameter $\beta$, unstable phase appears for smaller values of $%
\beta $, and therefore nonlinearity brings in the instability.
\end{abstract}
\maketitle

\section{Introduction}

The best-known theory of gravity in four dimensions is Einstein's general
relativity which is the most successful theory of gravity in describing our
universe at middle and large scale. A century after the fundamental
predictions of Einstein, the recent detection of gravitational waves is a
confirmation of this theory. However, we do not expect Einstein's theory to
remain valid at very high energies close to the Planck scale and therefore
the modification of general relativity is unavoidable. As we know, string
theory \cite{String1} and brane cosmology \cite{Brane} makes strong
predictions about the existence of extra dimensions and therefore among the
large variety of possible gravitational modifications, generalizing the
field equations in higher dimensions seems to be worthwhile. Lovelock
introduced a theory that modifies the Einstein's theory with terms keeping
the order of the field equations down to second order in derivatives in
higher dimensions \cite{Lovelock}. The resulting terms are free of ghost and
keep the generality of general relativity in four dimensions. It is worth to
mention that the second order Lovelock term which is known as the
Gauss-Bonnet term appears in the low energy effective action of string
theory \cite{String2}.

If one drops the necessity of the constancy of curvature of the horizon in
higher dimensions, there are many more possibilities for black hole
solutions in Lovelock gravity. This is due to the fact that Riemann tensor
appears in the field equation of Lovelock gravity. But, in Einstein gravity,
if one replaces the general $(n-2)$-dimensional space of positive constant
curvature with an $(n-2)$-dimensional space with positive curvature, it does
not alter the black hole potential. As an example of nonconstant curvaure
metric, one may use the infinite family of inhomogeneous metrics with
positive scalar curvature on products of spheres constructed by Bohm \cite%
{Bohm} or Einstein metric \cite{Einmetric}. The physical applications of Bohm
and Einstein metrics are studied in \cite{Hartnoll}. In \cite{Canfora} the
metric with nontrivial behavior that represents black hole of Lovelock-BI
gravity is found in even dimensions by allowing the base manifold to be
non-Einstein. Using the nonconstant curvature spaces as the
horizon of black holes in Lovelock gravity, the presence of the higher-order
gravity terms restricts the geometry of the boundary by imposing constraints
on its Weyl tensor \cite{Dotti}. These constraints bring new parameters in
the metric function and modify the properties of the black holes. After
Dotti and Gleiser who obtained an exact vacuum black hole solution in
Einstein-Gauss-Bonnet gravity \cite{Dotti}, the properties of such solutions
have been investigated in \cite{Tronc, Maeda, Bogdanos, Dadh}. The
spacetimes with Einstein manifold are investigated in third order Lovelock
gravity and it is shown that Weyl curvature must obey two kinds of algebraic
conditions \cite{Farhang1}. For black holes with nonconstant
curvature base manifolds, general tensorial conditions imposing on the
horizons by Lovelock field equations of an arbitrary order is obtained in
\cite{Ray}. Furthermore, it is found in \cite{Ray}, that these conditions
are equivalent to the ones in
terms of tensors involving the conformal Weyl tensors. Also, Birkhoff's
theorem is extended for such base manifolds using an
elementary method. The properties of such black holes in vacuum are
investigated in \cite{Ohashi}.

Our aim in this paper is to construct solutions of third order Lovelock
gravity with nonconstant curvature horizon in the presence of a nonlinear
electromagnetic field and investigate their properties. As we mentioned, the
nonlinearity of gravitational field equation with respect to Riemann tensor
has some effects on the properties of black holes with nonconstant
curvature. So, it is worth to investigate the effects of nonlinearity of
electromagnetic field on the properties of these kinds of black holes. The
properties of black holes with nonconstant curvature horizon in the
presence of Maxwell field have been investigated in \cite{Farhang2}. Here,
we want to investigate the effects of nonlinearity of electromagnetic field
on the properties of these solutions. Indeed, the existence of some
limitations in the Maxwell theory and the fact that the nonlinear
electrodynamics is richer than the linear Maxwell theory motivate one to
consider nonlinear electrodynamics. The kind of nonlinear electromagnetic
field which we consider is Born-Infeld (BI) electromagnetic field. Born and
Infeld proposed a specific model of nonlinear electrodynamics with the aim
of well behavior of the self-energy of a pointlike charge and avoiding
physical quantities to become infinite \cite{BI1}. The BI model was inspired
mainly to remedy the fact that the standard picture of a point particle
possesses an infinite self-energy, by placing an upper limit on the electric
field strength and considering a finite electron radius. The coupling of
nonlinear electrodynamics to gravity became of interest soon after that, and
the first solution of the Einstein equations for a pointlike BI charge was
obtained in \cite{Hoffmann}. After that Einstein-BI black holes were
revisited in \cite{EBI1, EBI2, EBI3, EBI4, EBI5, EBI6, EBI7}. Also, the
effects of nonlinearity of Born-Infeld (BI) electromagnetic field have been
investigated on the black hole solutions of Gauss-Bonnet \cite{GBBI} and
Lovelock gravities \cite{LovBI}. All of these black hole solutions in the
presence of the BI field have maximally symmetric horizons.
Also, the thermodynamics of these black holes with constant curvature horizons have been
studied so far \cite{LTherm, LTherm1, LTherm2, LTherm3, LTherm4}.
In this paper, we are supposed to consider a more general class of Einstein
spaces as the horizon, calculate the thermodynamic quantities and perform
stability analysis for such solutions.

The plan of the paper is organized as follows. In Sec. II we briefly review
the field equations of BI nonlinear electrodynamics coupled to Lovelock
gravity. Also, the structure of nonconstant curvature spaces with constant
Ricci scalar will be reviewed. In Sec. III higher dimensional BI black holes in Lovelock
gravity with special constraints on their horizons are derived and main the
properties of these solutions are discussed. Section IV dedicates to thermodynamics of the
solutions and stability is discussed by calculating the respective
quantities. Finally, we close the paper with a concluding section summarizing the
results.

\section{The Theory}

Born-Infeld Lagrangian leads to field equations whose spherically symmetric
static solution gives a finite value $\beta $ for the electrostatic field at
the origin. The constant $\beta $ appears in the BI Lagrangian as a new
universal constant. We begin with the action of Lovelock
gravity in the presence of nonlinear BI electromagnetic field in $D$
dimensions, which is written as
\begin{equation}
I=\int_{\mathcal{M}}d^{D}x\sqrt{-g}\left( -2\Lambda +\sum_{p=1}^{\bar{p}}\alpha
_{p}\mathcal{L}^{(p)}+\mathcal{L}(F)\right) ,\text{ \ \ \ \ }\mathcal{L}%
^{(p)}=\frac{1}{2^{p}}\delta _{\rho _{1}...\rho _{p}\sigma _{1}...\sigma
_{p}}^{\mu _{1}...\mu _{p}\nu _{1}...\nu _{p}}R_{\mu _{1}\nu _{1}}^{\text{ \
\ }\rho _{1}\sigma _{1}}...R_{\mu _{p}\nu _{p}}^{\text{ \ \ }\rho _{p}\sigma
_{p}}\text{\ \ \ \ \ \ \ \ \ \ \ }  \label{Act}
\end{equation}%
where $\Lambda $ is the cosmological constant, $\alpha _{p}$'s are the
Lovelock coupling constants with the choose of $\alpha _{1}=1$, and delta
symbol is a totally antisymmetrized product of Kronecker delta functions. In
this relation $\mathcal{L}(F)$ is the BI Lagrangian defined as%
\begin{equation}
\mathcal{L}(F)=4\beta ^{2}(1-\sqrt{1+\frac{F^{\mu \nu }F_{\mu \nu }}{2\beta
^{2}}}),  \label{BILag}
\end{equation}%
where $\beta $\ is the BI parameter and $F_{\mu \nu }:=\partial _{\mu
}A_{\nu }-\partial _{\nu }A_{\mu }$, with $A_{\mu }$ being the vector
potential. The relation (\ref{BILag}) reduces to the standard Maxwell form
$\mathcal{L}(F)=-F^{2}$, in the limit $\beta \rightarrow \infty $\ $\ $while
$\mathcal{L}(F)\rightarrow 0$ as $\beta \rightarrow 0$.

\bigskip The gravitational and electromagnetic field equations derived from
the action (\ref{Act}) read%
\begin{eqnarray}
\mathcal{G}_{\nu }^{\mu } &:&=\sqrt{g}\overset{\overline{p}}{\underset{p=0}{%
\sum }}\alpha _{p}\mathcal{G}_{\nu }^{(p)\mu }=\text{ }\frac{1}{2}g_{\mu \nu
}\mathcal{L}(F)+\frac{2F_{\mu \lambda }F_{\nu }^{\lambda }}{\sqrt{1+\frac{%
F^{2}}{2\beta ^{2}}}},\text{ }  \notag \\
\text{\ }\mathcal{G}_{\nu}^{(p)\mu} &=&-\frac{1}{2^{p+1}}\delta _{\rho \rho
_{1}\sigma _{1}...\rho _{p}\sigma _{p}}^{\mu \mu _{1}\nu _{1}...\mu _{p}\nu
_{p}}R_{\mu _{1}\nu _{1}}^{\text{ \ \ }\rho _{1}\sigma _{1}}...R_{\mu
_{p}\nu _{p}}^{\text{ \ \ }\rho _{p}\sigma _{p}}\text{\ \ \ }  \label{Geqm}
\end{eqnarray}

and

\begin{equation}
\nabla _{\nu }(\frac{\sqrt{-g}F^{\mu \nu }}{\sqrt{1+\frac{F^{2}}{2\beta ^{2}}%
}})=0.\text{\ \ }  \label{gradF}
\end{equation}

The kinds of spacetime we are interested in, have metrics of the form
\begin{equation}
ds^{2}=-f(r)dt^{2}+\frac{1}{f(r)}dr^{2}+r^{2}\gamma _{ij}(z)dz^{i}dz^{j},
\label{metric1}
\end{equation}%
that is a warped product of a $2$-dimensional \textit{Lorentzian}
submanifold $\mathcal{M}^{2}$ and a $(D-2)$-dimensional submanifold $%
\mathcal{K}^{(D-2)}$. We shall use alphabets , $i,$ $j,$ $m,.$.. to denote
space indices on the $(D-2)$-dimensional base manifold. Here we assume the
submanifold $\mathcal{K}^{(D-2)}$ with the unit metric $\gamma _{ij}$ to be
an Einstein manifold with nonconstant curvature but having a constant Ricci
scalar being%
\begin{equation}
\widetilde{R}=\kappa (D-2)(D-3),\text{ \ }  \label{Ricciscalar}
\end{equation}%
\ with $\kappa $ being the sectional curvature. Hereafter we use tilde for
the tensor components of the submanifold $\mathcal{K}^{(D-2)}.$ The Ricci
and Riemann tensors of the Einstein manifold are

\begin{eqnarray}
\text{\ \ \ \ \ \ \ }\widetilde{R}_{ij} &=&\kappa (D-3)\gamma _{ij},
\label{Ricci Ten} \\
\widetilde{{R}}{_{ij}}^{kl} &=&\widetilde{{C}}{_{ij}}^{kl}+\kappa ({\delta
_{i}}^{k}{\delta _{j}}^{l}-{\delta _{i}}^{l}{\delta _{j}}^{k})\text{\ },
\label{Riemm Ten}
\end{eqnarray}%
where $\widetilde{{C}}{_{ij}}^{kl}$ is the Weyl tensor of $\mathcal{K}%
^{(D-2)}$. In four dimensions, the Weyl tensor is zero, and Eq.\ (\ref{Riemm
Ten}) is satisfied for constant curvature manifolds. However, for dimensions
more than four, constant curvature manifolds are just special cases of
Einstein manifolds.

In \cite{Ray} the author shows that for an Einstein base manifold satisfying
Eq. (\ref{Riemm Ten}) the following constraints are imposed on the Weyl
tensor of the base manifold
\begin{equation}
\frac{1}{2^{n+1}}\delta
_{jp_{1}q_{1}...p_{n}l_{n}}^{il_{1}m_{1}...l_{n}m_{n}}\widehat{C}%
_{l_{1}m_{1}}^{p_{1}q_{1}}...\widehat{C}_{l_{n}m_{n}}^{p_{n}q_{n}}=\frac{%
(D-3)!\delta _{j}^{i}}{2(D-2p-3)!}\overset{p}{\underset{n=0}{\sum }}\binom{p%
}{n}(-\zeta _{1})^{n}\zeta _{p-n},  \label{Constn}
\end{equation}%
where the constants $\zeta _{p}$'s generally depend on the geometry of the
base manifold.

\section{Black Hole Solutions}

In this section, we want to introduce black hole solutions of third order
Lovelock gravity in the presence of BI field. For $p=2$ and $p=3$, Eq. (\ref%
{Constn}) may be written as
\begin{equation}
{\tilde{C}_{ki}}^{nl}{\tilde{C}_{nl}}^{kj}=\frac{1}{D}{\delta _{i}}^{j}{%
\tilde{C}_{km}}^{pq}{\tilde{C}_{pq}}^{km}\equiv \eta _{2}{\delta _{i}}^{j},
\label{eta2}
\end{equation}%
\begin{eqnarray}
&&2(4{\tilde{C}^{nm}}_{pk}{\tilde{C}^{kl}}_{ni}{\tilde{C}^{pj}}_{ml}+{\tilde{%
C}^{pm}}_{in}\tilde{C}^{jnkl}\tilde{C}_{klpm})  \notag \\
&=&\frac{2}{D}{\delta _{i}}^{j}\left( 4{\tilde{C}^{qm}}_{pk}{\tilde{C}^{kl}}%
_{qr}{\tilde{C}^{pr}}_{ml}+{\tilde{C}^{pm}}_{qr}\tilde{C}^{qrkl}\tilde{C}%
_{klpm}\right)   \notag \\
&\equiv &\eta _{3}{\delta _{i}}^{j}.  \label{eta3}
\end{eqnarray}%
These two conditions are first introduced in \cite{Dotti} and \cite{Farhang1}
respectively. Choosing $\bar{p}=3$ in the field equation (\ref{Geqm}),
making use of Eqs. (\ref{eta2}) and (\ref{eta3}), and the expressions in warped
geometry, the $\mathcal{G}_{t}^{t}$ component of the field equation could be
written as%
\begin{eqnarray}
0 &=&n\widehat{\alpha }_{0}r^{n-1}+\{[\widehat{\alpha }_{3}\psi ^{3}+%
\widehat{\alpha }_{2}\psi ^{2}+\left( 1+\frac{3\widehat{\alpha }_{3}\widehat{%
\eta }_{2}}{r^{4}}\right) \psi ]r^{n}\}%
{\acute{}}%
+(n-4)\widehat{\alpha }_{2}\widehat{\eta }_{2}r^{n-5}+(n-6)\widehat{\alpha }%
_{3}\widehat{\eta }_{3}r^{n-7}  \notag \\
&&+\frac{4\beta ^{2}r^{n-1}}{(n-1)}(1-\frac{1}{\sqrt{1+\frac{F^{2}}{2\beta
^{2}}}}).  \label{Gtt}
\end{eqnarray}%
where $n=D-1$, $\widehat{\alpha }_{p}$ and $\widehat{\eta }_{p}$ are defined
as $\widehat{\alpha }_{0}\equiv -2\Lambda /n(n-1)$, $\widehat{\alpha }%
_{p}\equiv (n-2)!\alpha _{p}/(n-2p)!$ and $\widehat{\eta }_{p}\equiv
(n-2p-1)!\eta _{p}/(n-1)!$\ for simplicity. In Eq. (\ref{Gtt}), the function
$\psi (r)$ is defined as%
\begin{equation}
\psi (r)\equiv \frac{\kappa -f(r)}{r^{2}}.  \label{Psi}
\end{equation}
We consider $\widehat{\alpha }_{2}$ and $\widehat{\alpha }_{3}$ as positive
parameters. It is also notable to mention that $\widehat{\eta }_{2}$ is
always positive, but $\widehat{\eta }_{3}$ can be positive or negative
relating to the metric of the spacetime. For the static spacetime (\ref%
{metric1}), Eq. (\ref{gradF}) can be satisfied by setting%
\begin{equation}
F^{rt}=\frac{\sqrt{(n-1)(n-2)}\beta q}{\sqrt{2\beta
^{2}r^{2(n-1)}+(n-1)(n-2)q^{2}}},  \label{EMField}
\end{equation}%
as the only nonvanishing component of $F^{\mu \nu }.$ A suitable vector
potential satisfying Eq. (\ref{gradF}) is
\begin{eqnarray}
A_{\mu } &=&-\sqrt{\frac{(n-1)}{2(n-2)}}\frac{q}{r^{(n-2)}}\digamma
(z)\delta _{\mu }^{0}, \\
\digamma (z) &=&_{2}F_{1}\left( \left[ \frac{1}{2},\frac{(n-2)}{2n-2}\right]
,\left[ \frac{3n-4}{2n-2}\right] ,-z\right) ,\text{ \ \ \ \ \ \ }z=\frac{%
(n-1)(n-2)q^{2}}{2\beta ^{2}r^{2n-2}}.
\end{eqnarray}%
Here $\digamma (z)$ is a hypergeometric function and appears in the
integration equation%
\begin{equation}
_{2}F_{1}\left( \left[ \frac{1}{2},b\right] ,\left[ b+1\right] ,-z\right)
=b\int \frac{t^{b-1}dt}{\sqrt{1+zt}}.
\end{equation}%
In the limit $\beta \rightarrow \infty $ $(z\rightarrow 0),$ the
hypergeometric function $\digamma (z)\rightarrow 1,$ thus $A_{\mu }$ will be
the gauge potential of the Maxwell field. Now by substituting Eq. (\ref%
{EMField}) in gravitational Eq. (\ref{Gtt}), after integrating one obtains
\begin{eqnarray}
0 &=&\widehat{\alpha }_{3}\psi ^{3}+\widehat{\alpha }_{2}\psi ^{2}+\left( 1+%
\frac{3\widehat{\alpha }_{3}\widehat{\eta }_{2}}{r^{4}}\right) \psi +%
\widehat{\alpha }_{0}+\frac{\widehat{\alpha }_{2}\widehat{\eta }_{2}}{r^{4}}+%
\frac{\widehat{\alpha }_{3}\widehat{\eta }_{3}}{r^{6}}-\frac{m}{r^{n}}
\notag \\
&&+\frac{4\beta ^{2}}{n(n-1)}\left[ 1-\sqrt{1+z}+\frac{(n-1)z}{(n-2)}%
\digamma (z)\right] .  \label{Eq3}
\end{eqnarray}%
In this relation, $m$ is the integration constant which is known as the
geometric mass and related to the ADM mass of the black hole. One of the real
solutions to this cubic equation may be written as
\begin{eqnarray}
\psi (r) &=&-\frac{\alpha _{2}r^{2}}{3\widehat{\alpha }_{3}}\left\{ 1-\left(
j(r)\pm \sqrt{\gamma +j^{2}(r)}\right) ^{1/3}+\gamma ^{1/3}\left( j(r)\pm
\sqrt{\gamma +j^{2}(r)}\right) ^{-1/3}\right\} ,  \notag \\
j(r) &=&-1+\frac{9\widehat{\alpha }_{3}}{2\widehat{\alpha }_{2}^{2}}-\frac{27%
\widehat{\alpha }_{3}^{2}}{2\alpha _{2}^{3}}\left( \widehat{\alpha }_{0}-%
\frac{m}{r^{n}}+\frac{\widehat{\alpha }_{3}\widehat{\eta }_{3}}{r^{6}}+\frac{%
4\beta ^{2}}{n(n-1)}\left[ 1-\sqrt{1+z}+\frac{(n-1)z}{(n-2)}\digamma (z)%
\right] \right) ,\text{ \ }  \notag \\
\text{\ \ \ \ }\gamma &=&\left( -1+\frac{3\widehat{\alpha }_{3}}{\widehat{%
\alpha }_{2}^{2}}+\frac{9\widehat{\alpha }_{3}^{2}\widehat{\eta }_{2}}{%
\widehat{\alpha }_{2}^{2}r^{4}}\right) ^{3}.  \label{fstat}
\end{eqnarray}%
One may note that constants $\widehat{\eta }_{2}$ and $\widehat{\eta }_{3}$
are evaluating on the $(n-1)$-dimensional boundary. In order to have the
effects of nonconstancy of the curvature of the horizon in the solutions,
we consider spacetimes with the dimension more than seven. As we expect when
$\widehat{\eta }_{2}=\widehat{\eta }_{3}=0$, the solution (\ref{fstat})
reduces to the solution of third order Lovelock gravity with constant
curvature horizon, in the presence of BI electromagnetic field \cite{LovBI}.

We can find the behavior of the metric for large $r$, using the fact that $%
_{2}F_{1}\left( [a,b];[c],-z\right) \rightarrow 1-\frac{ab}{c}z,$ and has a
convergent series expansion for $|z|<1$. Using definition (\ref{Psi})\ in
Eq. (\ref{Eq3}), one obtains
\begin{eqnarray}
0 &=&\hat{\alpha}_{0}r^{6}+(\kappa -f_{Lr})r^{4}+\hat{\alpha}_{2}[(\kappa
-f_{Lr})^{2}+\hat{\eta}_{2}]r^{2}+\hat{\alpha}_{3}[(\kappa -f_{Lr})^{3}+3%
\hat{\eta}_{2}(k-f_{Lr})+\hat{\eta}_{3}]  \notag \\
&&-\frac{m}{r^{n-6}}+\frac{q^{2}}{r^{2n-2}}-\frac{(n-1)^{2}(n-2)^{2}}{n(3n-4)\beta ^{2}}\frac{%
q^{4}}{r^{4n-4}},  \label{finf}
\end{eqnarray}%
where $f_{Lr}$ is the value of $f(r)$ at large values of $r$. The last term
in (\ref{finf}) is the leading BI correction to the electric charged black
hole in the large values of $r$ or $\beta $. One can see that the terms
including $q$ and $\beta $ vanish for very large values of $r$, and thus the
behavior of metric function is the same as that of third order Lovelock
gravity in vacuum and the asymptotic AdS solution may exist if Eq. (\ref%
{finf}) has positive real roots \cite{Farhang1, Myers}.

The Kretschmann scalar $R_{\mu \nu \rho \sigma }R^{\mu \nu \rho \sigma }$
diverges at $r=0$. Hence, there is an essential singularity located at $r=0$%
. More interesting is the behavior of the metric function close to origin
which reveals the variety of singular structures of the black hole
solutions. Using definition (\ref{Psi}) and the expansion of $%
_{2}F_{1}\left( \left[ a,b\right] ,\left[ c\right] ,-z\right) $ for large $z$%
, we can write Eq. (\ref{Eq3}) as
\begin{eqnarray}
0 &=&\hat{\alpha}_{3}(\kappa -f_{Sr})^{3}+\hat{\alpha}_{2}r^{2}(\kappa
-f_{Sr})^{2}+[3\hat{\alpha}_{3}\hat{\eta}_{2}+r^{4}](\kappa -f_{Sr})+(\hat{%
\alpha}_{0}+\frac{4\beta ^{2}}{n(n-1)})r^{6} \notag\\
&&+\hat{\alpha}_{2}r^{2}\hat{\eta}_{2}+\hat{\alpha}_{3}\hat{\eta}_{3}
-Cr^{n+5}-\frac{m-A}{r^{n-6}}-\frac{B}{r^{n-7}},\label{Near Origin}
\end{eqnarray}%
where $f_{Sr}$ is the value of  $f$  for small values of $r$ with $A,$ $B$
and $C$ being the constants defined as
\begin{eqnarray}
A &=&\frac{2(n-1)q^{2}}{n\sqrt{\pi }}\{\frac{2\beta ^{2}}{(n-1)(n-2)q^{2}}%
\}^{\frac{n-2}{2n-2}}\Gamma \lbrack \frac{3n-4}{2n-2}]\Gamma \lbrack \frac{1%
}{2n-2}], \\
B &=&\frac{2\beta q}{n}\sqrt{\frac{2(n-2)}{(n-1)}}\{1-\frac{(n-1)}{(n-2)}%
\frac{\Gamma \lbrack \frac{3n-4}{2n-2}]\Gamma \lbrack \frac{-1}{2n-2}]}{%
\Gamma \lbrack \frac{n-2}{2n-2}]\Gamma \lbrack \frac{2n-3}{2n-2}]}\}, \\
C &=&\frac{2\beta ^{3}}{n(n-1)(n-2)}\sqrt{\frac{2(n-2)}{(n-1)}}\{1+\frac{%
(n-1)}{(n-2)(2n-1)q}\frac{\Gamma \lbrack \frac{3n-4}{2n-2}]\Gamma \lbrack
\frac{-1}{2n-2}]}{\Gamma \lbrack \frac{n-2}{2n-2}]\Gamma \lbrack \frac{2n-3}{%
2n-2}]}\}.
\end{eqnarray}%
To find the behavior of the metric function $f$ near the origin $r=0,$ we
should find the solutions of the cubic equation below
\begin{eqnarray}
(\kappa -f_{Sr})^{3}+3\hat{\eta}_{2}(\kappa -f_{Sr})+\hat{\eta}_{3}-\frac{m-A}{\hat{\alpha}%
_{3}r^{n-6}}-\frac{B}{\hat{\alpha}_{3}r^{n-7}} &=&0.
\end{eqnarray}
For the solutions with nonconstant curvature horizon, the nature of singularity depends on the term
including $(m-A)$. For $m>A$, the
metric function approaches $-\infty $ as $r$ goes to zero and therefore the singularity is
spacelike. In this case the behavior of the solution resembles that of the
uncharged solution of third order Lovelock theory. While for $m<A$ the
solution resembles the charged solution in the presence of Maxwell field
having a timelike singularity. As it is seen for $n=7$ and $m=A$,
the metric function has a finite value at the origin which can be positive,
negative or zero depending on the parameters of the solution.

\begin{figure}[tbp]
\centering {\includegraphics[width=7cm]{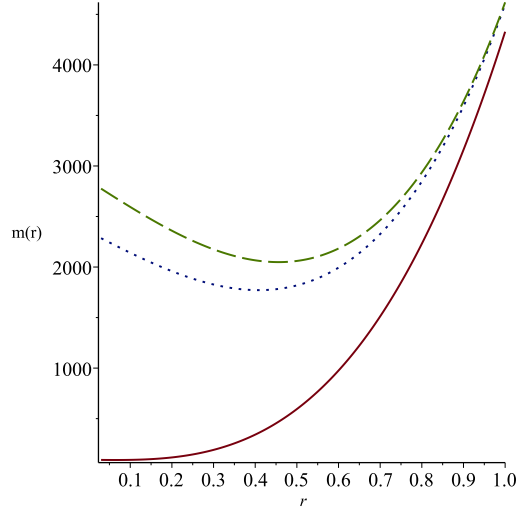}}
\caption{$m(r)$ versus $r$ for $n=9$, $\hat{\protect\alpha}_{0}=0.5$, $\hat{%
\protect\alpha}_{2}=2$, $\hat{\protect\alpha}_{3}=5$, $\hat{\protect\eta}%
_{2}=0.5$, $\hat{\protect\eta}_{3}=0.006$, $q=20$, and $\protect\beta =1$
(solid line), $\protect\beta =20$ (dotted line) and $\protect\beta =60$
(dashed line).}
\label{fig1}
\end{figure}

We could write the mass parameter $m$ in terms of horizon radius $r_{+}$ as
\begin{eqnarray}
m &=&\widehat{\alpha }_{0}r_{+}^{n}+\kappa r_{+}^{n-2}+\widehat{\alpha }%
_{2}[\kappa ^{2}+\widehat{\eta }_{2}]r_{+}^{n-4}+\widehat{\alpha }%
_{3}[\kappa ^{3}+3\widehat{\eta }_{2}\kappa +\widehat{\eta }_{3}]r_{+}^{n-6}
\notag \\
&&+\frac{4\beta ^{2}r_{+}^{n}}{n(n-1)}\left[ 1-\sqrt{1+z_{+}}+\frac{(n-1)z_{+}}{%
(n-2)}\digamma (z_{+})\right] ,
\end{eqnarray}%
where $z_{+}$ is the value of $z$ at $r=r_{+}$. To see how the value of mass
parameter characterize the nature of the horizon, we plot the mass parameter
as a function of the horizon radius for different values of $\beta $ which
are presented with solid, dotted, and dashed lines in Fig. \ref{fig1}. As it
is seen, two horizons exist for the dotted and dashed lines for certain
choice of $m$. If $m$ decreases, two horizons meet and black hole is
extreme. We call the value of mass parameter $m_{ext}$ in this case. This
condition happens when $r_{+}$ satisfies the following equation

\begin{eqnarray}
0 &=&n\widehat{\alpha }_{0}r_{+}^{n-1}+(n-2)\kappa r_{+}^{n-3}+(n-4)\widehat{%
\alpha }_{2}(\widehat{\eta }_{2}+\kappa ^{2})r_{+}^{n-5}+(n-6)\widehat{%
\alpha }_{3}(\widehat{\eta }_{3}+3\widehat{\eta }_{2}\kappa +\kappa
^{3})r_{+}^{n-7}  \notag \\
&&+\frac{4\beta ^{2}r_{+}^{n-1}}{(n-1)}(1-\sqrt{1+z_{+}}).  \label{ExtBH}
\end{eqnarray}%
This equation could not be solved analytically, but we just notice that the
black hole has two horizons for $m>m_{ext},$ and possesses a naked
singularity for $m<m_{ext}.$ For the solid line, one horizon exists for any
value of $m$. This means that in this case Eq. (\ref{ExtBH}) has no solution.

\section{Black hole Thermodynamics}

\label{Stab} The Hawking temperature of the black hole could be calculated
from the relation $T=(1/4\pi )(df/dr)_{r=r_{+}}$ where $r_{+}$ is the radius
of the outer horizon. Substituting in this relation we obtain the
temperature to be

\begin{eqnarray}
T &=&\frac{1}{4\pi r_{+}[r_{+}^{4}+2\kappa \widehat{\alpha }_{2}r_{+}^{2}+3%
\widehat{\alpha }_{3}(\widehat{\eta }_{2}+\kappa ^{2})]}\{nr_{+}^{6}\widehat{%
\alpha }_{0}+(n-2)\kappa r_{+}^{4}+(n-4)\widehat{\alpha }_{2}(\widehat{\eta }%
_{2}+\kappa ^{2})r_{+}^{2}  \notag \\
&&+(n-6)\widehat{\alpha }_{3}(\widehat{\eta }_{3}+3\kappa \widehat{\eta }%
_{2}+\kappa ^{3})+\frac{4\beta ^{2}}{(n-1)}r_{+}^{6}(1-\sqrt{1+z_{+}})\},
\label{Temp}
\end{eqnarray}%
where $r_{+}$ is the radius of the outer horizon.

In higher curvature gravity the area law of entropy, which states that the
black hole entropy equals one-quarter of the horizon area is
not satisfied. To calculate the entropy on the Killing horizon, we make use
of Wald prescription which is applicable for any black hole solution of
which the event horizon is a killing horizon \cite{Wald}. This is given by
the following integral on $(n-1)$-dimensional spacelike bifurcation surface

\begin{equation}
S=-2\pi \oint d^{n-1}x\sqrt{h}Y,\text{ \ \ \ \ \ }Y=Y^{abcd}\widehat{%
\varepsilon }_{ab}\widehat{\varepsilon }_{cd},\text{\ \ \ \ \ \ }Y^{abcd}=%
\frac{\partial \mathcal{L}}{\partial R_{abcd}}  \label{entropy}
\end{equation}%
in which $\mathcal{L}$ is the Lagrangian and $\widehat{\varepsilon }_{ab}$
is the binormal to the horizon. Straightforward calculations lead to the
following expression for the entropy on the horizon as
\begin{equation}
S=\frac{\Sigma _{\kappa }(n-1)r_{+}^{n-1}}{4}\left\{ \frac{1}{(n-1)}+\frac{%
2\kappa \widehat{\alpha }_{2}}{r_{+}^{2}(n-3)}+\frac{3\widehat{\alpha }%
_{3}(\kappa ^{2}+\widehat{\eta }_{2})}{r_{+}^{4}(n-5)}\right\} .
\label{Entro}
\end{equation}%
where $\Sigma _{\kappa }$ represents the volume of nonconstant-curvature
hypersurface. The first term in this expression is proportional to the area
of the horizon. It is seen that topological invariants also contribute to
the whole entropy of Lovelock black holes. The terms including $\widehat{%
\alpha }_{2}$ and $\widehat{\alpha }_{3}$ are present for the maximally
symmetric horizons, while the term including $\widehat{\eta }_{2}$
represents contribution coming from the Einstein horizon.

Comparing the field equation at large values of $r$ (\ref{finf}) with the equation of motion of
third order Lovelock equation for constant curvature horizon, one can find
that the ADM mass of the black hole is
\begin{eqnarray}
M &=&\frac{\Sigma _{\kappa }(n-1)m}{16\pi }=\frac{(n-1)\Sigma _{\kappa }}{%
16\pi }\{\widehat{\alpha }_{0}r_{+}^{n}+\kappa r_{+}^{n-2}+\widehat{\alpha }%
_{2}[\kappa ^{2}+\widehat{\eta }_{2}]r_{+}^{n-4}+\widehat{\alpha }%
_{3}[\kappa ^{3}+3\widehat{\eta }_{2}\kappa +\widehat{\eta }_{3}]r_{+}^{n-6}
\notag \\
&&+\frac{4\beta ^{2}r_{+}^{n}}{n(n-1)}\left[ 1-\sqrt{1+z_{+}}+\frac{%
(n-1)z_{+}}{(n-2)}\digamma (z_{+})\right] \}.  \label{ADM}
\end{eqnarray}

Note that from the Hawking temperature (\ref{Temp}), entropy (\ref{Entro})
or the mass parameter (\ref{ADM}), one can see that the case of $\beta =0$
or $q=0$ reduces to the case of uncharged Lovelock black hole with
nonconstant curvature horizon as expected \cite{Farhang1}. While in the case
of $\beta \rightarrow \infty $ the expressions reduce to those of charged
solution in the presence of Maxwell field. The electric field $E(r)$ is
defined by the relation $E(r)=-\Phi ^{^{\prime }}(r),$ in which $\Phi (r)$
is the electric potential and is derived by integrating the electric field.
For BI electromanetics the electric field is calculated to be%
\begin{equation}
E(r)=\frac{q}{\sqrt{\frac{q^{2}}{\beta ^{2}}+r^{2n-2}}},
\end{equation}%
from which we calculate electric potential measured at infinity with respect
to the horizon is

\begin{equation}
\Phi =\sqrt{\frac{n-1}{2n-4}}\frac{q}{r_{+}^{n-2}}\digamma (z_{+}).
\end{equation}

Also $Q$ which is called thermodynamic electric charge is related to the
charge via%
\begin{equation}
Q=\frac{q\Sigma _{\kappa }}{4\pi }\sqrt{\frac{(n-1)(n-2)}{2}}.
\end{equation}

To show that the solutions we obtained above, follow the first law of
thermodynamics, with the help of the following relation
\begin{equation}
\frac{d}{dr}(\frac{_{2}F_{1}\left( \left[ \frac{1}{2},b\right] ,\left[ b+1%
\right] ,-z\right) }{(n-3)r^{n-3}})=\frac{-1}{\sqrt{\frac{q^{2}}{\beta ^{2}}%
+r^{2n-4}}},
\end{equation}%
and making use of thermodynamic quantities that we derived, the equation%
\begin{equation*}
dM=T\partial {S+\Phi \partial Q,}
\end{equation*}
is easily satisfied.

\begin{figure}[tbp]
\centering {\includegraphics[width=7cm]{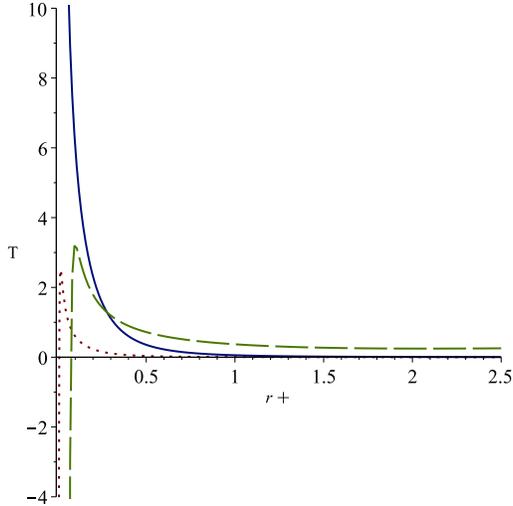}}
\caption{$T$ versus $r_{+}$ for $n=9$, $\hat{\protect\alpha}_{0}=0.1$, $\hat{%
\protect\alpha}_{2}=2$, $\hat{\protect\alpha}_{3}=5$, $\hat{\protect\eta}%
_{2}=1$ and $\hat{\protect\eta}_{3}=5$, for uncharged solution (solid line),
charged solution with $q=1$ (dashed line) and BI solution with $q=1$ and $%
\protect\beta =0.5$ (dotted line).}
\label{fig2}
\end{figure}

\begin{figure}[tbp]
\centering {\includegraphics[width=7cm]{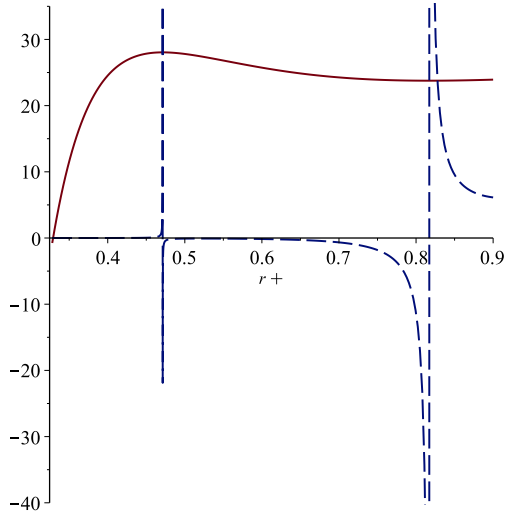}}
\caption{$T$ (solid line) and $C_{Q}$ (dashed line) versus $r_{+}$ for $n=8$%
, $\protect\kappa =1$, $\hat{\protect\alpha}_{0}=0.1$, $\hat{\protect\alpha}%
_{2}=1$, $\hat{\protect\alpha}_{3}=1/3$, $\hat{\protect\eta}_{2}=1$, $\hat{%
\protect\eta}_{3}=5$, $q=2$, and $\protect\beta =5$.}
\label{fig3}
\end{figure}

\begin{figure}[tbp]
\centering {\includegraphics[width=7cm]{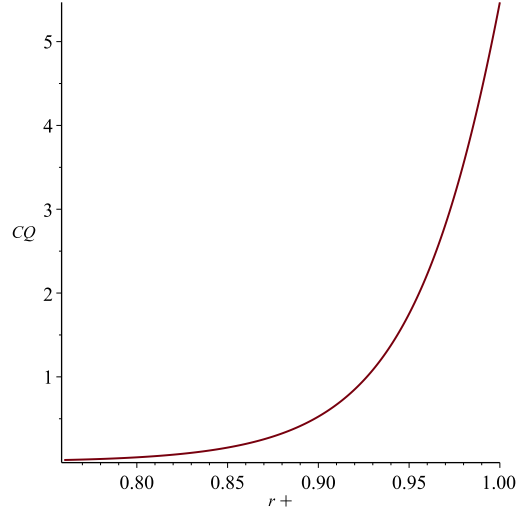}}
\caption{$C_{Q}$ versus $r_{+}$ for $n=9$, $\hat{\protect\alpha}_{0}=0.1$, $%
\hat{\protect\alpha}_{2}=1$, $\hat{\protect\alpha}_{3}=1/3$, $\hat{\protect%
\eta}_{2}=1$, $\hat{\protect\eta}_{3}=5$, and $q=0.1$.}
\label{fig4}
\end{figure}

\begin{figure}[tbp]
\centering {\includegraphics[width=7cm]{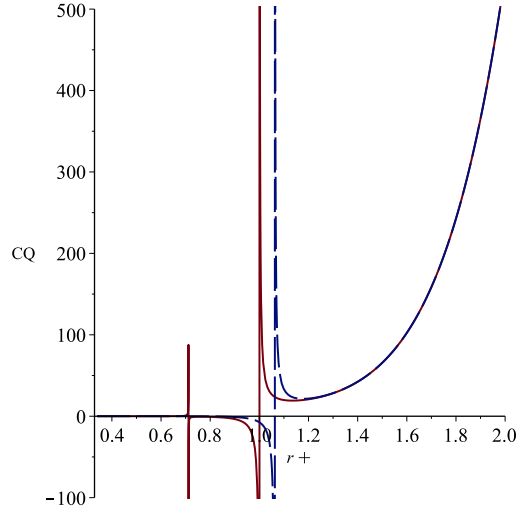}}
\caption{$C_{Q}$ versus $r_{+}$ for $n=9$, $\hat{\protect\alpha}_{0}=0.1$, $%
\hat{\protect\alpha}_{2}=1$, $\hat{\protect\alpha}_{3}=1/3$, $\hat{\protect%
\eta}_{2}=1$, $\hat{\protect\eta}_{3}=5$, $q=0.1$, and $\protect\beta =0.3$
(solid line), $\protect\beta =0.1$ (dashed line) .}
\label{fig5}
\end{figure}

Now we are ready to study the influence of the nonlinearity of the
electromagnetic field on the existence of the thermal stability of the black
hole solutions. The method of performing stability of black holes of Einstein
gravity may be found in \cite{Hawking, Kastor}. As we are
investigating the stability in the canonical ensemble, the charge is fixed
and the heat capacity is defined by the relation
\begin{equation}
C_{Q}=(\frac{\partial M}{\partial T})_{Q}=T(\frac{\partial S}{\partial T}%
)_{Q}  \label{Hc}
\end{equation}%
An increase in temperature for fixed charge will result in an increase in
the entropy leading local stability. Thus, positive heat capacity implies
that the black hole is locally stable. The relation for $C_{Q}$ is
complicated and we do not write it here. Instead we follow a numerical
analysis. We display $C_{Q}-r_{+}$ and $T-r_{+}$ diagrams. To see the effect
of the BI term on the thermodynamic of the system, first we plot temperature
versus the black hole horizon $r_{+}$ in Fig. \ref{fig2} for uncharged,
electric charged and BI black holes. For a range of values of $q$ and $\beta
$, the temperature of BI black holes has a maximum and a local minimum
values at $r_{+}=r_{1}$ and $r_{+}=r_{2}$ respectively for which $\frac{dT}{%
dr_{+}}=0$. It is zero at $r_{+}=r_{0}$ for which extreme black hole can
exist, where the value of $r_{0\text{ }}$ is getting closer to zero for BI
black holes.
Due to the existence of the term including $\beta ,$ and $\widehat{\eta }%
_{3} $ that could be negative, the Hawking temperature given by the relation
(\ref{Temp}), could be negative which is unphysical. So we depict capacity
as a function of $r_{+}$ in the region that temperature is
positive. Considering relation (\ref{Hc}), one can see that the heat capacity
is zero at $r_{+}=r_{0}$ and blows up at $r_{+}=r_{1}$ and $r_{2}$, so the
black hole has a phase transition at these points. The graphs of $T$ and $%
C_{Q}$ vs. $r_{+}$ are shown for positive $\kappa $ in Fig. \ref{fig3}. It
is seen that for positive $\kappa $ small black holes ($r_{0}<$ $r_{+}<r_{1}$%
) and large ones ($r_{+}>r_{2}$) are stable while there exists an
intermediate unstable phase with horizon area $\ r_{1}<r_{+}<r_{2}$. This
case is similar to Einstein-BI and Lovelock-BI black hole with
constant-curvature horizon \cite{EBI1, LovBI}.
But the case is different for $\kappa =0$. It is known that for Lovelock
black holes with constant curvature horizon and $\kappa =0$, the Lovelock
parameters do not appear in relation of temperature, entropy and heat
capacity. Thus Lovelock correction has no contribution in the heat capacity
and therefore Lovelock-BI black hole are locally stable in the whole range
of $r_{+}$ \cite{EBI5}. But for our new solution, with nontrivial boundary,
the existence of Lovelock coefficients in addition to the parameters that
appear due to the nonconstancy of the horizon makes drastic changes to the
relations. The entropy is no longer proportional to the area. For $\kappa =0$
black holes with nonconstant-curvature horizon, an unstable phase exists
similar to what we explained for solutions with positive $\kappa $. To see
the effect of the nonlinearity of the BI field on the stability of the black
hole, first we display $C_{Q}$ versus $r_{+}$ for charged Lovelock Black
hole with nonconstant-curvature horizon in the presence of the Maxwell field
in Fig. \ref{fig4}. We see that for a chosen value of $q$ the black hole
is stable in the whole range of $r_{+}.$ In Fig. \ref{fig5} the heat
capacity is depicted for the BI solution with the same fixed value of $q$
but for different values of $\beta $. The interesting result is that the
unstable phase appears when $\beta $ is decreased. This means that
nonlinearity of the field creates instability of the black hole.

\section{Concluding Remarks}

New solutions of Lovelock theory in the presence of BI field have been
investigated. The horizon space consumed is nonmaximally symmetric Einstein
space which has nonzero Weyl curvature. The supplementary conditions on the
Weyl tensor, have a nontrivial contribution in the solution in terms of
chargelike parameters. The behavior of the solutions has been presented at
infinity which shows that asymptotic behavior of the solution is the same as
that of uncharged solution and charged solution in the presence of Maxwell
field. Thus, the matter field has no contribution in the metric function at
infinity. Near the origin, the behavior of the solution is more interesting
and more variety exists for the nature of the singularity of the black
holes. For the special value of $m$ in eight dimensions ($n=7$) the metric
function has a finite value at the origin which can be positive, negative or zero.
For dimensions higher than seven ($n>6$)
depending on the values of the parameters of the solution, the singularity
could be spacelike resembling the solution without a matter field, or
timelike which is the behaviour of singularity of the electric charged
solution in the presence of the Maxwell field. We
also showed that these kinds of black holes could have one or two horizons
or possess naked singularity depending on the parameters of the solution.
Next, we calculated thermodynamical quantities in order to investigate the
stability of the black holes. Plotting temperature versus horizon radius for
solutions in vacuum and in the presence of the Maxwell and BI field
separately, we found that as $\beta $ increases, one may have smaller
extreme black holes. By calculating the heat capacity and applying numerical
analysis for positive $\kappa $, we found that small and large black hole
are locally stable, while there exists an intermediate unstable phase. This
is similar to the BI black holes with constant curvature horizon. Lovelock
parameters do not appear in the relation of temperature, entropy and heat
capacity of Lovelock black holes with flat horizon, and therefore Lovelock
correction has no contribution in these variables. Thus, these kind of
Lovelock-BI black hole are locally stable in the whole range of $r_{+}$ for
any value of $\beta $. But, for our solutions with nonconstant curvature
with $\kappa =0$, the appearance of the parameters in the thermodynamic
quantities makes drastic changes in the properties of black holes in such a
way that one may have unstable phase. To check the effect of the
nonlinearity of the BI field, we compared the plot of heat capacity verses
horizon for charged solution in the presence of Maxwell field for a fixed $q$,
and then in the presence of a BI field for that fixed value of $q$ but for
different values of $\beta$. The result indicates that while the black hole
is stable in the whole range of $r_{+}$ in the presence of Maxwell field, or
either BI field with large $\beta $, instability appears for smaller values
of $\beta $, and therefore nonlinearity brings in instability.


\begin{thebibliography}{99}
\bibitem{String1} B. A. Campbell, M. J. Duncan, N. Kaloper, and K. A. Olive,
Nucl. Phys. B\textbf{351}, 778 (1991); I. Antoniadis, J. Rizos, and K.
Tamvakis, Nucl. Phys. B\textbf{415}, 497 (1994).

\bibitem{Brane} L. Randall, R. Sundrum, Phys. Rev. Lett. \textbf{83}, 3370 (1999);
\textbf{83}, 4960 (1999); G. Dvali, G.
Gabadadze, and M. Porrati, Phys. Lett. B \textbf{485}, 208 (2000); G. Dvali
and G. Gabadadze, Phys. Rev. D \textbf{63}, 065007 (2001).

\bibitem{Lovelock} D. Lovelock, J. Math. Phys. \textbf{12}, 498 (1971).

\bibitem{String2} B. Zwiebach, Phys. Lett. B \textbf{156}, 315 (1985); D. J.
Gross and J. H. Sloan, Nucl. Phys. B\textbf{291}, 41 (1987).

\bibitem{Bohm} C. Bohm, Invent. Math. 134, 145 (1998).

\bibitem{Einmetric} H. Lu, D. N. Page, and C. N. Pope, Phys. Lett. B \textbf{%
593}, 218 (2004); J. P. Gauntlett, D. Martelli, J. F. Sparks, and D. Waldram,
Adv. Theor. Math. Phys. \textbf{8}, 987 (2004).

\bibitem{Hartnoll} G.W. Gibbons, S.A. Hartnoll, and C.N. Pope, Phys. Rev. D
\textbf{67}, 084024 (2003).

\bibitem{Canfora} F. Canfora and A. Giacomini, Phys. Rev. D \textbf{82}, 024022 (2010).

\bibitem{Dotti} G. Dotti and R.J. Gleiser, Phys. Lett. B \textbf{627}, 174 (2005).

\bibitem{Tronc} G. Dotti, J. Oliva, and R. Troncoso, Phys. Rev. D \textbf{75}, 024002 (2007);
G. Dotti, J. Oliva, and R. Troncoso, Int. J. Mod. Phys. A
\textbf{24}, 1690 (2009); G. Dotti, J. Oliva, and R. Troncoso, Phys. Rev. D
\textbf{82}, 024002 (2010).

\bibitem{Maeda} H. Maeda, Phys. Rev. D \textbf{81}, 124007 (2010); H. Maeda,
M. Hassaine, and C. Martinez, J. High Energy Phys. 08 (2010) 123.

\bibitem{Bogdanos} C. Bogdanos, C. Charmousis, B. Gouteraux, and R. Zegers,
J. High Energy Phys. 10 (2009) 037.

\bibitem{Dadh} J. M. Pons and N. Dadhich, Eur. Phys. J. C \textbf{75}, 280 (2015).

\bibitem{Farhang1} N. Farhangkhah and M. H. Dehghani, Phys. Rev. D \textbf{90}, 044014 (2014).

\bibitem{Ray} S. Ray, Classical Quantum Gravity \textbf{32}, 195022 (2015).

\bibitem{Ohashi} S. Ohashi and M. Nozawa, Phys. Rev. D \textbf{92}, 064020 (2015).

\bibitem{Farhang2} N. Farhangkhah, Int. J. Mod. Phys. D \textbf{25}, 1650030 (2016).

\bibitem{BI1} M. Born and L. Infeld , Nature (London) \textbf{132}, 1004 (1933).

\bibitem{Hoffmann} B. Hoffmann, Phys. Rev. D \textbf{47}, 877 (1935).

\bibitem{EBI1} A. Garcia, H. Salazar, and J.F. Plebanski,
Nuovo. Cimento. Soc. Ital. Fis. A \textbf{84}, 65 (1984);
G. W. Gibbons and D. A. Rasheed, Nucl. Phys. B\textbf{454}, 185 (1995).

\bibitem{EBI2} H.P. de Oliveira, Classical Quantum Gravity \textbf{11}, 1469 (1994);
E. Ayon-Beato and A. Garcia, Phys. Rev. Lett. \textbf{80}, 5056 (1998).

\bibitem{EBI3} Sh. Fernando, Phys.Rev. D \textbf{74}, 104032 (2006).

\bibitem{EBI4} T. Tamaki and T. Torii, Phys. Rev. D \textbf{62}, 061501 (2000);
G. Clement and D. Gal'tsov, Phys. Rev. D \textbf{62}, 124013 (2000).

\bibitem{EBI5} R. G. Cai, D. W. Pang, and A. Wang, Phys. Rev. D \textbf{70}, 124034 (2004).

\bibitem{EBI6} M. Aiello, R. Ferraro, and G. Giribet, Phys. Rev. D \textbf{70}, 104014 (2004).

\bibitem{EBI7} T. K. Dey, Phys. Lett. B \textbf{595}, 484 (2004).

\bibitem{GBBI} M. H. Dehghani and S. H. Hendi, Int. J. Mod. Phys. D \textbf{%
16}, 1829 (2007); D. Zou, Z. Yang, R. Yue, and P. Li, Mod. Phys. Lett. A
\textbf{26}, 515 (2011).

\bibitem{LovBI} M. Aiello, R. Ferraro, and G. Giribet, Phys. Rev. D \textbf{70%
}, 104014 (2004); M. H. Dehghani, S. H. Hendi, A. Sheykhi, and H. Rastegar
Sedehi, J. Cosmol. Astropart Phys. 02 (2007) 020; M. H. Dehghani, N. Alinejadi, and S.
H. Hendi, Phys. Rev. D \textbf{77}, 104025 (2008).

\bibitem{LTherm} R. Banerjee and D. Roychowdhury, Phys. Rev. D \textbf{85}, 044040 (2012).

\bibitem{LTherm1} J. X. Mo and W. B. Liu, Eur. Phys. J. C \textbf{74}, 2836 (2014).

\bibitem{LTherm2} Y. S. Myung, Y. W. Kim, and Y. J. Park, Phys. Rev. D \textbf{78}, 084002 (2008).

\bibitem{LTherm3} O. Miskovic and R. Olea, Phys. Rev. D \textbf{77}, 124048 (2008);
O. Miskovic and R. Olea, Phys. Rev. D \textbf{83}, 064017 (2011); D.
C. Zou, Z. Y. Yang, R. H. Yue, and P. Li, Mod. Phys. Lett. A \textbf{26}, 515 (2011).

\bibitem{LTherm4} P. Li, R. H. Yue, and D. C. Zou, Commun. Theor. Phys.
\textbf{56}, 845 (2011).

\bibitem{Myers} R. C. Myers and B. Robinson, J. High Energy Phys. 08 (2010) 067.

\bibitem{Hawking} S.W. Hawking and D. N. Page, Commun. Math. Phys. \textbf{87}, 577 (1983).

\bibitem{Wald} R. M. Wald, Phys. Rev. D \textbf{48}, R3427 (1993); V. Iyer and
R. M. Wald, Phys. Rev. D \textbf{50}, 846 (1994); T. Jacobson, G. Kang, and R.
C. Myers, Phys. Rev. D \textbf{49}, 6587 (1994).

\bibitem{Kastor} D. Kastor, S. Ray, and J. Traschen, Classical Quantum Gravity \textbf{26}, 195011 (2009).
\end{thebibliography}
\end{document}